\documentclass[superscriptaddress, reprint]{revtex4-2}
\usepackage{commath}
\newcommand{\be}{\begin{equation}}
	
	\newcommand{\ee}{\end{equation}}
\newcommand{\bea}{\begin{eqnarray}}
	\newcommand{\eea}{\end{eqnarray}}
\newcommand{\ba}{\begin{array}}
	\newcommand{\ea}{\end{array}}

\newcommand{\se}{Schr\"{o}dinger equation}
\newcommand{\bl}{\begin{flalign}}
	\newcommand{\enl}{\end{flalign}}

\newcommand{\pa}{\partial}

\newcommand{\tdse}{time-dependent Schr\"{o}dinger equation\ }

\newcommand{\half}{\frac{1}{2}}

\usepackage{siunitx}
\usepackage{graphicx}
\usepackage{amssymb}
\usepackage{epstopdf}
\usepackage{amsmath,dsfont}
\usepackage{bm}

\usepackage{hyperref, xcolor}
\usepackage{physics}
\usepackage{braket}

\newcommand{\eq}[1]{Eq.~\eqref{#1}}
\newcommand{\Eq}[1]{Equation~\eqref{#1}}
\newcommand{\fig}[1]{Fig.~\ref{#1}}

\newcommand*{\rom}[1]{\expandafter\@slowromancap\romannumeral #1@}
\renewcommand{\bf}{\mathbf}
\newcommand{\mc}{\mathcal}
\renewcommand{\Re}{\operatorname{Re}}
\renewcommand{\Im}{\operatorname{Im}}

\newcommand{\proj}[1]{\ket{#1}\bra{#1}}
\newcommand{\bs}{\begin{split}}
\newcommand{\es}{\end{split}}

\renewcommand{\bf}{\mathbf}
\renewcommand{\Re}{\operatorname{Re}}
\renewcommand{\Im}{\operatorname{Im}}

\usepackage[capitalize]{cleveref}

\begin{document}

	\title{ Topological Quantum Molecular Dynamics}
	\author{Yujuan Xie}

	\affiliation{Department of Chemistry and Department of Physics, Westlake University, Hangzhou, Zhejiang 310030, China}
	\affiliation{Institute of Natural Sciences, Westlake Institute for Advanced Study, Hangzhou 310024, China.}
	
	\author{Ruoxi Liu}
	\affiliation{Department of Chemistry and Department of Physics, Westlake University, Hangzhou, Zhejiang 310030, China}
	
	\author{Bing Gu}
	\email{gubing@westlake.edu.cn}
	\affiliation{Department of Chemistry and Department of Physics, Westlake University, Hangzhou, Zhejiang 310030, China}
	\affiliation{Institute of Natural Sciences, Westlake Institute for Advanced Study, Hangzhou 310024, China.}

\begin{abstract}
We develop a unified quantum geometric framework to understand reactive quantum dynamics. The critical roles of the quantum geometry of adiabatic electronic states in both adiabatic and non-adiabatic quantum dynamics are unveiled. A numerically exact, divergence-free topological quantum molecular dynamics method is developed through a discrete local trivialization of the projected electronic Hilbert space bundle over the nuclear configuration space. In this approach, the singular electronic quantum geometric tensor— Abelian for adiabatic dynamics and non-Abelian for non-adiabatic dynamics—is fully encoded in the global electronic overlap matrix. With numerical illustrations, it is  demonstrated that atomic motion—whether adiabatic or non-adiabatic—is governed not only by the variation in electronic energies with nuclear configurations (potential energy surface) but also by the variation in electronic states (electronic quantum geometry).

\end{abstract}
	
	\maketitle

\section{Introduction}

The Born-Oppenheimer approximation is the cornerstone of modern chemistry \cite{born1927, tannor2007}. It assumes a separation of timescales between electronic and nuclear motion, arising from the disparity in their masses. Under this approximation, electronic motion adjusts instantaneously to the nuclear geometry, while the nuclei move along the adiabatic ground state potential energy surfaces (APES). Consequently, molecular motion including vibrational dynamics and chemical reaction rate constants are entirely determined by the topography of the APES, particularly the stationary points and transition states. This framework serves as the foundation for our understanding of ground-state chemistry \cite{tannor2007}.

In the conventional Born-Oppenheimer (BO) picture, only electronic energies, that can be determined from first principles by the solving the electronic \se\ (known as electronic structure or quantum chemistry calculations \cite{szabo1996, thijssen2007}), are  relevant, while the electronic states themselves do not play any  role. Apart from the adiabatic approximation (e.g using a single APES), the Born-Oppenheimer approximation invokes two additional assumptions: (i) the neglect of diagonal Born-Oppenheimer corrections, which modify the APESs, and (ii) the neglect of geometric phase effects, assuming that the gauge connection can be made to  vanish globally. Improving upon Born-Oppenheimer dynamics remains a challenge, even in the adiabatic regime.
The diagonal Born-Oppenheimer corrections diverge at conical intersections (CIs), and are not even integrable \cite{meek2016}. Conical intersections, ubiquitous in polyatomic molecules, play critical roles in  a wide range of photochemical and photophysical processes \cite{larson2020, domcke2011, prigogine2002} 
, analogous to  the transition state in ground-state chemistry. 
  Furthermore, it is difficult to construct a vector potential from ab initio data, as it requires a single-valued gauge with complex-valued electronic wavefunctions, whereas the electronic wavefunction is real-valued in quantum chemistry. The vector potential is not only singular at CIs but also carries a gauge-dependent branch cut for each CI, analogous to the vector potential in a Dirac monopole \cite{heras2018}.

To describe nonadiabatic dynamics, one adopts the Born-Huang expansion of the molecular wavefunction, a natural generalization of the Born-Oppenheimer ansatz that includes relevant electronically excited states. In this Born-Huang framework, the first-derivative coupling, also known as the nonadiabatic coupling, is commonly used to account for nonadiabatic transitions despite the presence of second-derivative couplings.
In addition to the challenges associated with the diagonal Born-Oppenheimer potential and vector potentials inherited from the Born-Oppenheimer ansatz, the equations of motion using the Born-Huang ansatz are, however, plagued by divergences in derivative couplings. Typically, approximate treatments, such as quasi-diabatization and the  vibronic coupling model Hamiltonians, are employed. Besides the common Born-Oppenheimer-Huang framework, there are other frameworks for quantum molecular dynamics, such as quantum trajectory methods based on the hydrodynamic de Broglie-Bohm formulation of quantum mechanics \cite{bohm1952, garashchuk2003, lopreore1999, wyatt2005, gu2017a} and exact factorization \cite{abedi2012}, which usually cannot avoid divergences in derivative couplings and can even introduce new ones (e.g., the singularity of the quantum potential at wavefunction nodes).

Here, we unify the  adiabatic and nonadiabatic molecular quantum dynamics into  a quantum geometric framework. Realizing that the physical origin of the inevitable divergences in quantum molecular dynamics is the nontrivial topology of the molecular fiber bundle, we discretize the full molecular bundle through a discrete local trivialization procedure, using a finite set of electron-nuclear product spaces. It is shown that the global electronic overlap matrix, the overlap between many-electron states at different nuclear geometries, encodes the intrinsic quantum geometric structure of the projected molecular fiber bundles and accounts for all effects beyond Born-Oppenheimer, thus providing an exact framework for molecular quantum dynamics. This includes the diagonal Born-Oppenheimer corrections for each APES, the geometric (or topological) phase effects due to conical intersections, and electronic transitions for nonadiabatic dynamics. We show that the matrix elements between nearby nuclear geometries are directly connected to the electronic quantum geometric tensor, a measure of the quantum geometry of the molecular fiber bundle. The momentum space analog has  found widespread applications in describing band topology for condensed matter systems, where the base space is the Brillouin zone \cite{provost1980, ma2010}.

The molecular fiber bundle is formed by a finite set of electronic states serving as the fiber and the nuclear configuration space as the base space. The electronic quantum geometric tensor comprises two closely related components: the Riemannian geometry and the Berry geometry, corresponding to the real and imaginary parts of the electronic quantum geometric tensor, respectively. This tensor quantifies how electronic states change in response to variations in nuclear configurations.
The Riemannian geometry, characterized by the quantum metric (the real part of the electronic quantum geometric tensor), influences nuclear motion as a scalar potential. Hence, its effects are local. The Berry geometry \cite{berry1984}, characterized by the Berry curvature, affects nuclear motion through a vector potential in the Born-Oppenheimer framework \cite{mead1992, xie2019, mead1980, mead1980a}. Both effects, neglected in Born-Oppenheimer dynamics, become significant in the presence of electronic (quasi)degeneracies, particularly at conical intersections.
Electronic degeneracy, whether symmetry-allowed or accidental, has been shown to be ubiquitous in polyatomic molecules \cite{domcke2011, truhlar2003, prigogine2002}. The presence of such degenerate seams leads to a nontrivial topology of the molecular fiber bundle, and even when energetically inaccessible, it can strongly influence nuclear dynamics.
However, such effects go beyond the Born-Oppenheimer approximation and are extremely challenging to fully incorporate into the Born-Oppenheimer framework due to the singularities in the first- and second-derivative couplings and a singular branch cut in the vector potentials.  

Instead of using various corrections with seemingly different physical origins to gradually go beyond Born-Oppenheimer dynamics as in the conventional framework, we found that, surprisingly, the only missing term in Born-Oppenheimer dynamics is the electronic overlap matrix. 
 In our numerically exact topological molecular quantum dynamics, the APES accounts for the variation of electronic energies with nuclear configurations, while the electronic overlap matrix describes the variation of the corresponding electronic states, i.e. the topology of the molecular fiber bundle.

A robust, divergence-free, yet numerically exact method for molecular quantum dynamics that fully incorporates the influences of electronic quantum geometry into nuclear motion is developed. 
We show that the Riemannian geometry and Berry geometry are encoded, respectively, in the amplitude and phase of the global electronic overlap matrix. That is, the discretized electronic overlap matrix fully encapsulates the quantum geometric information of the adiabatic electronic states and directly relates to the Berry connection \cite{berry1984, mead1982} and the quantum geometric tensor \cite{provost1980, ma2010}.
While the Berry curvature and quantum metric become singular at electronic degeneracies, the overlap matrix remains bounded within the range of $[-1,1]$.
Specifically, when the electronic Hamiltonian is time-reversal symmetric, the Berry geometry arising from the presence of conical intersections becomes topological rather than geometrical. This means that it encodes global topological information about the electronic Hilbert space, leading to geometric phase effects even when these intersections are energetically inaccessible \cite{berry1984, longuet-higgins1958, pancharatnam1956, cohen2019, martinazzo2024, zhu2022, gu2024a, wittig2012a, manini2000, ryabinkin2017, babikov2005, yuan2018a, kendrick1995, kendrick2002, domcke2011, yarkony2019, ryabinkin2014, aharonov1987}.
These effects are critical for understanding a wide range of chemical processes, ranging from vibrational spectra \cite{kendrick1995, babikov2005}, through nonadiabatic molecular dynamics \cite{zilberg2000}, such as in the photodissociation of the phenol molecule \cite{li2024, yuan2020, valahu2023, nix2008}, to ground-state chemical reactions like the elementary hydrogen exchange reaction \cite{yuan2018a}.
While the classical treatment of nuclei has provided useful insights into many chemical phenomena, a full quantum molecular dynamics treatment is necessary for understanding the underlying quantum geometry and is urgently desired given recent experimental advancements particularly in ultrafast spectroscopy  probing the quantum nature of nuclei, revealing phenomena such as passage through conical intersections, cold molecules, molecular interference, geometric phase effects, and electronic coherence \cite{scholes2017, rafiq2023, matselyukh2022a,halpin2014}.

We demonstrate that for adiabatic quantum dynamics, it is possible to incorporate geometric phase effects into nuclear motion using only ground-state information. All quantum geometric information is fully contained in the single intrastate electronic overlap matrix. By contrast, conventional approaches, such as quasi-diabatization and vector potential methods, require excited-state information and nonadiabatic couplings. Exact diabatization does not exist due to a topological obstruction, it requires the vanishing of the Berry curvature matrix, see \cref{app:dia} for details. The local diabatic representation provides a straightforward ab initio approach to incorporating geometric phase effects into nuclear quantum dynamics using only the information from a single potential energy surface, despite the presence of CIs.

The quantum geometric framework is straightforwardly generalized to  nonadiabatic molecular quantum dynamics by simply including electronically excited states in the overlap matrix, leading to a multi-state global electronic overlap matrix. By doing so, the Abelian electronic quantum geometric tensor becomes non-Abelian. Our method does not require the electronic wave function to be smooth across the entire configuration space, making it applicable under any gauge fixing. This allows for the direct use of electronic states obtained from electronic structure calculations, which typically carry random phases assigned in matrix diagonalization subroutines (i.e., random gauge fixings), without any postprocessing \cite{gu2023b, gu2024a, zhu2024}. Moreover, the precise location and energy of conical intersections are not essential for our calculations.

Our approach thus eliminates the difficulties associated with constructing vector potentials and locating the CI seam in multi-dimensional configuration space. The location and number of CIs are typically not known a priori for a given molecule. The utility of our method in describing geometric molecular quantum dynamics is demonstrated first through a  vibronic model, then through a realistic phenol photodissociation model with more complex potential energy surfaces, and finally through an ab initio modeling of H$_3^+$ combined with electronic structure calculations.

Atomic units are used throughout $e = \hbar = m_\text{e} = 1$.

\section{Adiabatic Quantum Dynamics}

We first consider adiabatic molecular dynamics, where only a single potential energy surface is involved. This encompasses most chemical reactions that are not triggered by electronic excitation.
The Born-Oppenheimer approximation begins by employing a product ansatz for the molecular wave function
\be 
\Psi(\bf r, \bf R, t) = \psi(\bf r; \bf R) \chi(\bf R, t), 
\ee
 where $\psi(\bf r; \bf R)$ is the ground state of the electronic Hamiltonian, i.e., 
 \be 
 H_\text{BO}(\bf R) \psi(\bf r; \bf R) = V(\bf R) \psi(\bf r; \bf R).
 \ee 
 The electronic Hamiltonian consists of the electron kinetic energy operators and the Coulomb interaction between all charged particles  
 \be 
 \begin{split}
 H_\text{BO}(\bf R)  =& \sum_{i=1}^n -\half \nabla ^2_i + \sum_{i < j} \frac{1}{\abs{\bf r_i - \bf r_j}} - \sum_{i, I} \frac{Z_I}{\abs{\bf r_i - \bf R_I}} \\
 &+ \sum_{I < J} \frac{Z_IZ_J}{\abs{\bf R_I - \bf R_J}} 
 \end{split}
 \ee
where $i (I)$  runs through all electrons (nuclei). It depends parametrically on the nuclear geometry $\bf R$ through the electron-nuclear Coulomb interaction. Further interactions such as spin-orbit couplings can be included in the electronic Hamiltonian. 

Under the Born-Oppenheimer approximation, the \tdse for the nuclear wave packet $\chi(\bf R, t)$ is given by
 
\be 
i \hbar \pd{\chi(\bf R, t)}{t} = \del{\hat{T}_\text{N} + V(\bf R)} \chi(\bf R, t)
\label{eq:120}
\ee  
where $V(\bf R)$ is the ground state APES determined by the electronic \se, $\hat{T}_\text{N}$ is the nuclear kinetic energy operator. Taking the classical limit, the nuclear motion reduces to classical molecular dynamics.
It is evident from \eq{eq:120} that the nuclear motion is completely determined by the landscape of the APES.

The Born-Oppenheimer ansatz incurs a gauge structure, originating from the fact that the electronic and nuclear wave packets are“redundant” description of the molecular wavefunction. Specifically, there is a local U(1) gauge freedom, meaning that the molecular wavefunction remains invariant under the gauge transformation
\be 
\begin{split}
\phi'(\bf r; \bf R) &= \phi(\bf r; \bf R) e^{-i \theta(\bf R)}, \\ 
\chi'(\bf R, t) &= e^{i \theta(\bf R)} \chi(\bf R, t)
\end{split}
\ee    
for any $\theta(\bf R)$. 
Thus, a gauge fixing is required.

It is implicitly assumed in the Born-Oppenheimer dynamics that the gauge can be fixed such that
\be
F_\mu(\bf R) \equiv \braket{\psi(\bf R) | \pa_\mu  \psi(\bf R)} = 0. 
\label{eq:pt}
\ee 
where $F_\mu(\bf R) $ is the gauge-dependent Berry connection, and $\pa_\mu \equiv \pdv{\bf R_\mu}$ is the nuclear derivative. \cref{eq:pt} is the so-called parallel transport gauge \cite{uhlmann1986, zhou2020a}.

\subsection{Discrete Local Trivialization}
We now show that how  the electronic quantum geometry of the adiabatic electronic manifold can arise in nuclear quantum dynamics and how it can be fully incorporated by the global electronic overlap matrix. To achieve this, we first discretize the molecular fiber bundle using a discrete variable locally diabatic ansatz for the electron-nuclear wave function
\be 
\Psi(\bf r, \bf R, t) = \sum_n C_n(t) \psi(\bf r; \bf R_n) \chi_n(\bf R)
\label{eq:121}
\ee

Here, the nuclear configuration space is discretized using a discrete variable representation (DVR) of the nuclear coordinate operators, $\mathbf{R}_\mu$, where $\mu = 1, 2, \dots, N$, and $\psi(\mathbf{r}; \mathbf{R}_n)$ represents the electronic eigenstates at molecular geometry $\mathbf{R}_n$. The discrete variable representation basis sets can be considered as the eigenstates of position operators (i.e., Dirac delta functions) projected onto the variational (computational) space, making them highly localized.
We have chosen this particular basis sets due to their locality and orthogonality. 
Owing to  locality, we can  choose the electronic states at the center of each nuclear basis as an electronic basis set. Since the electronic states do not {explicitly} depend on the nuclear configuration, each vibronic basis function $\psi(\mathbf{r}; \mathbf{R}_n) \chi_n(\mathbf{R})$ is a product state and is therefore topologically trivial. That is, the entire molecular bundle, which can be topologically nontrivial, is covered by a finite number of product spaces, a process known as local trivialization \cite{frankel2011}.
Inserting \eq{eq:121} into the \tdse yields 

\be 
i \dot{\bf C}\del{t} = \del{  \bf T \bf A + \bf V} \bf C(t)
\label{eq:main}
\ee 
where  
\be A_{mn} = \braket{\psi(\bf R_m) | \psi(\bf R_n)}_{\bf r}
\ee 
is the overlap matrix between the many-electron wave functions at two nuclear geometries. The absolute value of the overlap matrix represents the fidelity between two electronic states, whereas its phase encodes the geometric phase information. The overlap matrix reduces to an identity when two geometries coincide, i.e., $A_{nn} = 1$ due to normalization. If the same DVR basis set is used to describe the Born-Oppenheimer wave packet dynamics, the equation of motion becomes

\be 
i  \dot{\bf C}\del{t} = \del{ \bf T + \bf V} \bf C(t). 
\label{eq:bo}
\ee  

The equation of motion using the locally diabatic ansatz (\eq{eq:main}) is similar to the Born-Oppenheimer dynamics, with, however, a crucial difference being that the nuclear kinetic energy matrix is dressed by the electronic overlap matrix.
 The overlap matrix measures how electronic states vary with nuclear geometry. If we set 
\be A_{mn} = 1  ~~~\forall m, n, \ee
\eq{eq:main} reduces precisely to the Born-Oppenheimer dynamics \eq{eq:bo}. Physically, this corresponds to a trivial quantum geometry where the electronic state does not vary with nuclear configuration. Therefore, the Born-Oppenheimer approximation, from a quantum geometric perspective,  correspond to trivializing the topology of the molecular fiber bundle.
This provides an alternative view for understanding the Born-Oppenheimer approximation, whereas the conventional understanding, based on the mass difference between electrons and nuclei, is obscured by divergences in the vibronic couplings at the electronic degenerate manifold, as shown below.

In the continuum limit $n \rightarrow \infty$, \eq{eq:main} becomes an singular integro-differential equation ($C_n(t) \chi_n(\bf R) \rightarrow \chi(\bf R_n, t) \rightarrow \chi(\bf R, t)$)
\be 
i \pdv{\chi(\bf R, t)}{t} = \int \dif \bf R'  T(\bf R, \bf R')  \mc{A}(\bf R, \bf R') \chi(\bf R' ,t) + V(\bf R)  \chi(\bf R, t)
\label{eq:main2}  
\ee 
where $T(\bf R, \bf R') = \braket{\bf R| \hat{T}_\text{N}|\bf R'}$ is coordinate representation of the kinetic energy operator. Qualitatively speaking, \cref{eq:main} can be considered as a lattice regularization of \eq{eq:main2} using discrete variable representation.

\subsection{Electronic quantum geometry}
The discrete global electronic overlap matrix fully encodes the  quantum geometry of the projected electronic fiber bundle.  To see this, suppose that we have fixed the gauge freedom so that the electronic state is locally smooth around $\bf R_m$, consider a nearby geometry $\bf R_n = \bf R_m + \bf \Delta$, the overlap matrix reads 

\begin{align}
	\ln & \braket{\psi(\mathbf{R}_m) }{\psi(\mathbf{R}_m + \mathbf{\Delta})}_{\mathbf{r}} \nonumber \\
	& = \sum_\mu F_\mu \Delta_\mu - \sum_{\mu, \nu} \frac{1}{2} Q_{\mu \nu}(\mathbf{R}_m) \Delta_\mu \Delta_\nu + \mathcal{O}(\Delta^3)
	\label{eq:131}
\end{align}
where 
\be 
Q_{\mu \nu}(\bf R) = \mel{\pa_\mu \psi(\bf R)}{1 - \mc{P}(\bf R) }{\pa_\nu \psi(\bf R)}_{\bf r}
\ee 
is the gauge-invariant electronic quantum geometric tensor and $ \mc{P}(\bf R) \equiv \proj{\psi(\bf R)}$ is a projection operator.
Note that we have only retained terms up to second order in \eq{eq:131}. This is because, in the limit of $\Delta \rightarrow 0$, only the first and second order terms contribute to the nuclear motion, as the nuclear kinetic energy operator is a second-order differential operator, i.e., scaling as $\mc{O}(\Delta^{-2})$ upon discretization. We can thus conclude that the order at which the electronic quantum geometry will influence the nuclear quantum dynamics is determined by the kinetic energy operator of the nuclei.

The electronic quantum geometric tensor is reminiscent of the quantum geometric tensor of the Bloch band structure of crystals, which is useful in studying continuous quantum phase transitions \cite{zanardi2007, camposvenuti2007, gianfrate2020, provost1980, kolodrubetz2017}. The real symmetric part of the electronic quantum geometric tensor 
\be  g_{\mu \nu} = \Re Q_{\mu\nu} = \half \del{Q_{\mu \nu} + Q_{\nu \mu} }
\ee is the Riemannian metric tensor that measures the distance of the electronic states in the projective Hilbert space. As $\bf F$ is imaginary, \eq{eq:131} leads to $ - \ln \abs{  \braket{\psi(\bf R_m) }{ \psi(\bf R_m + \bf \Delta)}_{\bf r}  }^2 \approx  g_{\mu \nu}\Delta_\mu \Delta_\nu $. The imaginary antisymmetric part \be \Omega_{\mu \nu} = 2\Im Q_{\mu \nu} = i^{-1} \del{  Q_{\mu \nu} - Q_{\nu \mu} } = -\Omega_{\nu \mu} 
\ee is the Berry curvature. {Thus,  in addition to the electronic energies, the quantum geometry of the electronic states,  both the Riemannian geometry and the Berry geometry, influence molecular quantum dynamics. }

The electronic quantum geometry becomes important when the electronic states undergo substantial variation in the configuration space, i.e., when the overlap matrix deviates significantly from unity. If the adiabatic states remain unchanged with molecular geometry, i.e., $A_{mn} = 1$, \eq{eq:main} simplifies to the Born-Oppenheimer dynamics. However, regions where electronic states vary considerably with molecular geometry are common in polyatomic molecules, particularly, around avoided crossings, quasi-degeneracies, and conical intersections \cite{larson2020, domcke2011, yarkony2019}. In the case of conical intersections specifically, the electronic overlap matrix deviates sharply from unity \cite{zhu2024}.

There are two main reasons why the electronic quantum geometry does not appear in the Born-Oppenheimer dynamics. 
Firstly, in the Born-Oppenheimer dynamics, the assumption that a global gauge can be found where the real-valued electronic wave functions are single-valued and smooth (i.e., differentiable) such that the Berry connection vanishes is not valid e.g., in the presence of conical intersections. More generally, when the molecular fiber bundle is topologically nontrivial. The Berry connection (or vector potential) can be made locally the real-valued electronic wave function is double-valued, and thus cannot be made globally smooth \cite{domcke2011}. In other words, it is impossible to find the global parallel transport gauge.    
Imposing a single-valued gauge necessarily leads to complex-valued electronic wave functions, with the phase factor determined during the gauge-fixing process.
The second approximation involves neglecting the diagonal Born-Oppenheimer correction \cite{meek2016}, 
\be
E_\text{DBOC} = \mel{\psi(\bf R)  } {\sum_\mu -\frac{1}{2 M_\mu} \pa_\mu^2 }{\psi(\bf R)} 
\ee  which is contained in the quantum metric. Unlike the quantum metric, the diagonal Born-Oppenheimer correction is not gauge-invariant.

It is instructive to consider the exact adiabatic equation of motion using the Born-Oppenheimer ansatz without invoking further approximations, other than the electronic state truncation. Assuming that a single-valued gauge for the electronic states can be found, this yields \cite{hetenyi2023}
\be 
i \pd{}{t} \chi(\bf R, t) = \qty{
\sum_\mu -\frac{D_\mu^2 }{2M_\mu} + { V(\bf R) + G(\bf R)}} \chi(\bf R, t)    
\label{eq:adia}
\ee 
where $D_\mu \equiv {\pa_\mu + F_\mu(\bf R)}$ is the covariant derivative. \cref{eq:adia} provides a gauge-covariant description of nuclear quantum dynamics. It is valid in the single-valued gauge, where the electronic wave functions are complex-valued, and the gauge connection does not vanish. \eq{eq:adia} offers a gauge-covariant description of nuclear dynamics.

Hypothetically, it is possible to incorporate quantum geometry into Born-Oppenheimer dynamics by solving \eq{eq:adia}. However, this approach suffers from two severe challenges: singularities in the geometric tensor and random gauge fixings in ab initio electronic structures. The electronic quantum geometric tensor becomes singular at electronic degeneracy points, such as conical intersections. This occurs because, around a conical intersection, electronic states vary substantially even with small changes in configuration. This divergence is not a mathematical artifact but reflects the impossibility of finding a parallel transport gauge. Moreover, all electronic states obtained from electronic structure calculations carry either a random sign (in the $Z_2$ gauge for real-valued electronic wave functions under time-reversal symmetry) or a random phase (in the $U(1)$ gauge for complex-valued wave functions) due to gauge freedom, and there is no general recipe to find a global single-valued gauge in a high-dimensional space $d > 2$.

Both limitations associated with \eq{eq:adia} are eliminated in \eq{eq:main}, as it does not require a specific gauge fixing and is divergence-free since the overlap matrix is bounded within the range $[-1,1]$. These advantages arise from a crucial hidden difference between the Born-Oppenheimer ansatz and our local diabatic ansatz concerning gauge fixing. While the Born-Oppenheimer ansatz requires a global single-valued gauge, our ansatz imposes no such constraints by employing the global electronic overlap matrix to capture the intrinsic quantum geometry of the molecular fiber bundle. Even adiabatic electronic states with random gauge fixings can be directly used in \eq{eq:main}. Moreover, instead of describing Berry geometry through a gauge connection and quantum metric as a scalar potential, all the electronic geometry is captured within a single overlap matrix.

\section{Nonadiabatic molecular quantum dynamics}
The formalism discussed above can be easily extended to nonadiabatic molecular quantum dynamics, where electronic transitions among adiabatic electronic states play an important role. In the conventional approach to nonadiabatic dynamics, the Born-Oppenheimer ansatz is generalized to the Born-Huang expansion for the total molecular wave function
\be
\Psi(\bf r, \bf R, t) = \sum_{\alpha=1}^N \phi_\alpha(\bf r; \bf R)\chi_\alpha(\bf R, t) = \bm \phi \cdot \bm \chi
\label{eq:bh}
\ee 
where $\phi_\alpha(\bf r; \bf R)$ are the adiabatic electronic states, defined as the electronic eigenstates of the electronic Hamiltonian $H_\text{BO}(\bf R)$ at nuclear geometry $\bf R$. In ab initio treatments, the electronic Hamiltonian contains all the Coulomb interactions between charged particles and the electronic kinetic energy operators. With a finite number of electronic states, this expansion defines a projected sub-Hilbert space to solve molecular quantum dynamics. Our locally diabatic ansatz becomes \cite{gu2023b, gu2024a}
\be 
\Psi(\bf r, \bf R, t) = \sum_{\alpha=1}^N C_{n\alpha}(t) \phi_\alpha(\bf r; \bf R_n)\chi_n(\bf R, t) 
\ee

Inserting this ansatz into the time-dependent molecular \se\ yields the equation of motion for the expansion coefficients
\be 
i \dot{C}_{m\beta}(t) = V_{m \beta} C_{m\beta}(t) + \sum_{n, \alpha} T_{mn} A_{mn}^{\beta \alpha} C_{n\alpha}(t)
\label{eq:main3}
\ee 
where $V_{m\beta} = V_{\beta}(\bf R_m) $ is the electronic energy at $\bf R_m$. 

The electronic overlap matrix now carries the electronic state index 
\be 
A_{mn}^{\beta \alpha} = \braket{\phi_\beta(\bf R_m) }{ \phi_\alpha(\bf R_n)}_{\bf r}
\ee 

Trivializing the quantum geometry, i.e., assuming that all electronic states belonging to the same quantum number are equivalent and are orthogonal if the quantum numbers are different 
\be
A_{mn}^{\beta \alpha} \approx \delta_{\beta \alpha}, 
\ee 
\eq{eq:main3} reduces to the BO dynamics. From a chemical intuition, this approximation typically cannot hold because the adiabatic electronic states are ordered by energy, whereas the electronic character can drastically change when a bond breaks or forms or when the electronic states have charge transfer character.

Following the analysis in the adiabatic case, supposing a locally smooth gauge has been found around $\bf R_m$, let $\bf R_n = \bf R_n + \bf \Delta$, 
\be 
\ln \bf A_{mn} \approx \bf F_\mu \Delta_\mu - \half \bf Q_{\mu \nu} \Delta_\mu \Delta_\nu 
\ee  
where  \be 
[\bf F_\mu]_{\beta \alpha} 
\equiv 
F_\mu^{\beta \alpha}(\bf R) = \braket{\phi_\beta(\bf R) | \pa_\mu \phi_\alpha(\bf R)}
\ee 
is the nonadiabatic generalization of the  gauge connection. 
 Its diagonal elements are the Berry connections for all electronic states, the off-diagonal elements $\alpha \ne \beta$ are  known as the nonadiabatic coupling. 
Here  
\be 
Q^{\beta \alpha}_{\mu \nu} =  \mel{\pa_\mu \phi_\beta }{ 1 - \mc{P}(\bf R) }{ \pa_\nu \phi_\alpha},
\label{eq:gt}
\ee 
is the non-Abelian electronic quantum geometric tensor characterizing the geometric information of the electronic Hilbert space.  
It is gauge-invariant under the local U(1) gauge transformation $\phi'_\alpha(\bf r; \bf R) = e^{i\theta_\alpha(\bf R)} \phi_\alpha(\bf r; \bf R)$. 
In \eq{eq:gt}, $\mc{P}(\bf R) = \sum_{\alpha=1}^N \proj{\phi_\alpha(\bf R)}$ the electronic projection operator at nuclear geometry $\bf R$. The electronic geometric tensor vanishes in the complete electronic basis set limit. Intuitively, in the complete basis set limit (with necessarily infinity number of electronic states),  the electronic states are complete for any nuclear configuration, thus the projection operator $\mc{P}(\bf R) = 1$ does not depend on the nuclear configuration. The topology of the electron-nuclear fiber bundle is trivial.

In our picture, the nonadiabatic transitions arising from the first- and second-derivative couplings, the vector potential accounting for geometric phase effects, and the diagonal BO corrections are unified into a single overlap matrix. Nonadiabatic transitions occur because of the  electronic character similarity between ground and excited electronic wavefunctions between two nuclear configurations and are inherently nonlocal. Interestingly, this framework allows for long-range nonadiabatic transitions and suggests the possibility of direct electronic transitions between non-neighboring electronic states. Additionally, it naturally forbids transitions at trivial crossings.

We discuss the challenges in the conventional Born-Huang ansatz when fully incorporating electronic quantum geometry. Similar to the Born-Oppenheimer ansatz for adiabatic dynamics, a gauge fixing condition, which ensures that the adiabatic electronic states are globally smooth, is implicitly imposed in the Born-Huang expansion; otherwise, the nuclear kinetic energy operator cannot act on the electronic wave functions.
If such a parallel transport gauge can be established, the diagonal elements of the gauge connection matrix vanish, i.e., $F_\mu^{\alpha \alpha} = 0$. If the second-derivative coupling is further neglected, a widely adopted approximation, one arrives at a nonadiabatic dynamics framework where only the first-derivative coupling, $F^{\beta \alpha}_\mu$ ($\beta \ne \alpha$), induces nonadiabatic transitions. In this case, the geometric nature of the electronic states is lost.

To see how geometry can be incorporated, we assume that a single-valued gauge can be defined over the entire configuration space. Inserting the Born-Huang ansatz (\eq{eq:bh}) into the time-dependent molecular Schrödinger equation yields the equation of motion for the nuclear wavepackets (expressed in a gauge-covariant form to emphasize the geometric structure; see \cref{app:bh} for derivation details)
\be 
\sum_\mu -\frac{1}{2M_\mu} \del{\pa_\mu \bf I + \bf F_\mu}^2 \bm \chi + \del{\bf V + \bf G} \bm \chi  = {i \pd{}{t} } \bm \chi   
\label[type]{eq:111}
\ee 
where $\bm \chi(\bf R, t) = \sbr{\chi_1, \chi_2, \dots, \chi_N}$ is a column vector containing all nuclear wave functions, $\bf I$ is an $N \times N$ identity matrix in the electronic subspace, and $\bf V$ is a diagonal matrix whose diagonal elements correspond to the APESs. The index $\mu$ runs over all nuclear degrees of freedom, each associated with a mass $M_\mu$. 
Here the  scalar coupling 
\be  G_{\beta \alpha}(\bf R) = \sum_\mu  \frac{1}{2M_\mu} g^{\beta \alpha}_{\mu \mu} 
\ee 
arises from the quantum metric.

Recognizing $\bf D_\mu = \pa_\mu \bf I + \bf F_\mu$ as a matrix-valued covariant derivative, which generalizes the Abelian covariant derivative in \eq{eq:adia}, \eq{eq:main} provides a gauge-covariant description of nonadiabatic molecular quantum dynamics within the Born-Huang framework. However, this approach faces significant challenges in practical applications. Firstly, all electronic structure codes provide real electronic wave functions with a random gauge choice. Such random gauges can be fixed to the parallel transport gauge, but this results in double-valued electronic states in the presence of a CI. Finding an additional gauge transformation to eliminate this double-valuedness is highly challenging. Moreover, both the connection $\bf F_\mu$ and the electronic geometric tensor (EGT) $\bf G$ exhibit singularities at electronic degeneracies such as conical intersections. These singularities make \eq{eq:main} unsuitable for numerical simulations of nuclear wavepacket dynamics. All these challenges originate from the difficulty of defining a gauge in which the adiabatic electronic states remain globally smooth with respect to nuclear configurations.

\section{Generalization to Arbitrary Fiber Bundles}
{
	The discrete local trivialization procedure developped for electron-nuclear fiber bundles can be generalized to arbitrary  fiber bundles with nontrivial quantum geometry (topology), including non-differential ones. Fiber bundles are a general mathematical structure that is used to describe a wide range of physics and chemistry problems, including vibrational-rotational dynamics, nonadiabatic transitions, band structure, gauge theory, optics, and quantum electrodynamics.
	
	Consider a fiber bundle with a base space $\mc{M} \subset \mathbb{R}^d$ and a fiber $\mc{F}$, $d$ is the base space dimension. 
	The Hamiltonian, or more generally the generator for time translation, for the composite system can be generically partitioned  to 
	\be
	\mc{H} =  \mc{T}(\hat{P}) + H(x; X)
	\ee
	where $\mc{T}(\hat{P}) = \sum_{n \in \mathbb{Z}^+} t_n \hat{P}^n $ is the part containing the  momentum operator $\hat{P} = -i\nabla_{X}$ of the dynamical variables $X$ of the base space. For non-relativistic  variables, $\mc{T} \propto \hat{P}^2$. The fiber Hamiltonian $H(X_n)$, not necessarily Hermitian, consists of all terms describing the fiber degrees of freedom $x$ and their interactions with the base and any $X$-dependent terms representing the interactions between base variables. If the fiber is an open quantum system with dissipation and decoherence, we can simply replace the Hamiltonian with a Liouville superoperator $\mc{L}$ and use density matrices instead of states.
	
	If $H(X)$ is independent of $X$, the fiber bundle is a simple product space $\mc{M} \times \mc{F}$.
	Besides the case in which the quantum geometric tensor is singular due to degeneracy as in the case of conical intersections, it can also occur that the rank of $H(X)$ varies with $X$ (referred to as a pseudo-fiber bundle) so that the quantum geometric tensor is ill-defined. For instance, the non-Hermitian Hamiltonian at a exceptional point is defective. 
	The differential geometry language is conceptually appealing but is of limited utility for practical calculations due to the singularity and non-differentiable, especially for dimensionality of the base space $d \ge 2$ whereby it can be extremely difficult to find the singular hypersurfaces.
	
	In parallel to the treatment for molecular fiber bundles, we develop a discrete local trivialization procedure to describe the quantum dynamics on such fiber bundles. First, the base space is discretized by e.g. discrete variable representation $\set{X_n}$. 
	Upon a discretization, we can diagonalize the fiber Hamiltonian $H(X_n)$  
	\be H(X_n) \ket{\phi_{n\alpha}^\text{R}} =  {V}_{n \alpha} \ket{\phi_{n\alpha}^\text{R}} \ee
	to obtain the right ``adiabatic'' states $\ket{\phi_{n\alpha}^\text{R}}$ and complex eigenvalues $V_{n \alpha}$. For non-Hermitian Hamiltonian matrix, there is a set of biorthogonal states $\braket{\phi_{n\alpha}^\text{L} | \phi_{n\beta}^\text{R}} =  \delta_{\alpha\beta}$ for each $X_n$. 
	For Hermitian Hamiltonians, $\ket{\phi_{n\alpha}^\text{R}}^\dag = \bra{\phi_{n\alpha}^\text{L}}$ and the eigenvalues are real.
	
	The fiber bundle is described by a set of product states $\ket{n\alpha} \equiv \ket{\phi_{n\alpha}^\text{R}} \otimes \ket{X_n}$, $\ket{\Psi(t)} = \sum_{n\alpha} C_{n\alpha}(t) \ket{n\alpha}$. In this composite state space, the fiber Hamiltonian $H(X_n)$ is diagonal and the kinetic energy operators are represented by 
	\be \hat{P}^\alpha \rightarrow \braket{X_n| \hat{P}^\alpha | X_{n'}} \mc{A}_{nn'} \ee
	for any $\alpha \ge 1$. 
	Here $\mc{A}_{nn'}$ is the overlap matrix between the fibers at different base points with matrix elements 
	\be 
	\mc{A}_{nn'}^{\beta \alpha} = \braket{\phi_\beta^\text{L}(X_n)| \phi_{\alpha}^\text{R}(X_{n'})}. 
	\ee
	The matrix elements for momentum operators can  be easily computed  within the primitive basis set used for discretization. 
	Proceeding as for the molecular fiber bundles, supposing a gauge fixing that is locally smooth around $X_n$, we can apply a cumulant expansion of the overlap matrix $\mc{A}_{nn'}$ ($X_{n'} = X_n + \Delta$), 
	\be 
	\ln \mc{A}_{nn'} \approx \sum_{n=1}^N \sum_{\mu_1, \cdots, \mu_n=1}^{d}  \kappa_{\bm \mu}^{(n)} \Delta_{\mu_1} \Delta_{\mu_2} \cdots \Delta_{\mu_n}
	\ee
	The first order is the gauge connection  $\kappa^{(1)}_{ \mu}(X) = \braket{\phi_{\beta}^\text{L}(X)| \pa_\mu \phi_{\alpha}^\text{R}(X)}$, and the second order is the non-Hermitian non-Abelian quantum geometric tensor 
	\be 
	\mc{Q}_{\mu \nu}(X) = \sum_{\mu \nu} \braket{\pa_\mu \phi_{\alpha}^\text{L}(X) | 1 - \mc{P}_\text{NH}(X)| \pa_\nu \phi_{\alpha}^\text{R}(X)}
	\ee 
	where \be \mc{P}_\text{NH}(X) = \sum_{\alpha=1}^{\dim(\mc{F}(X_n))} \ket{\phi_{\alpha}^\text{L}(X)} \bra{\phi_{\alpha}^\text{R}(X)}
	\ee                        
	is the generalized projection operator for non-Hermitian systems, $\dim(H(X_n))$ is the dimension of the Hilbert space at $X_n$. 
	
	The fiber overlap  $\mc{A}_{nn'}$ provides a conceptually simple and practically useful way to encode the quantum geometry and topology over the quantum geometric tensor. It is well-defined and bounded for any (pseudo-)fiber bundle even when the second- and higher-order quantum geometric tensors cannot be defined.
	The equation of motion is given by 
	\be   
	i \dot{\bf C}(t) = \qty( \mc{T} \mc{A}  + \mc{V}) \bf C(t)\
	\label{eq:main3}
	\ee 
	where $\mc{T}$ is the matrix representation of the $\mc{T}(\hat{P})$, $\mc{V}$ is a diagonal matrix with elements correspond to the eigenvalues of the fiber Hamiltonian $H(X_n)$.
	\Eq{eq:main3} is a generalized form of \eq{eq:main} and \eq{eq:main2} for arbitrary fiber bundles. It provides a divergence-free approach to model the exact quantum dynamics on non-differential and non-Hermitian fiber bundles.	
}

\section{Numerical Illustrations and Discussion} \label{sec:discuss}

\subsection{Adiabatic dynamics}

\begin{figure*}[htbp]
	\centering
	\includegraphics[width=0.9\textwidth]{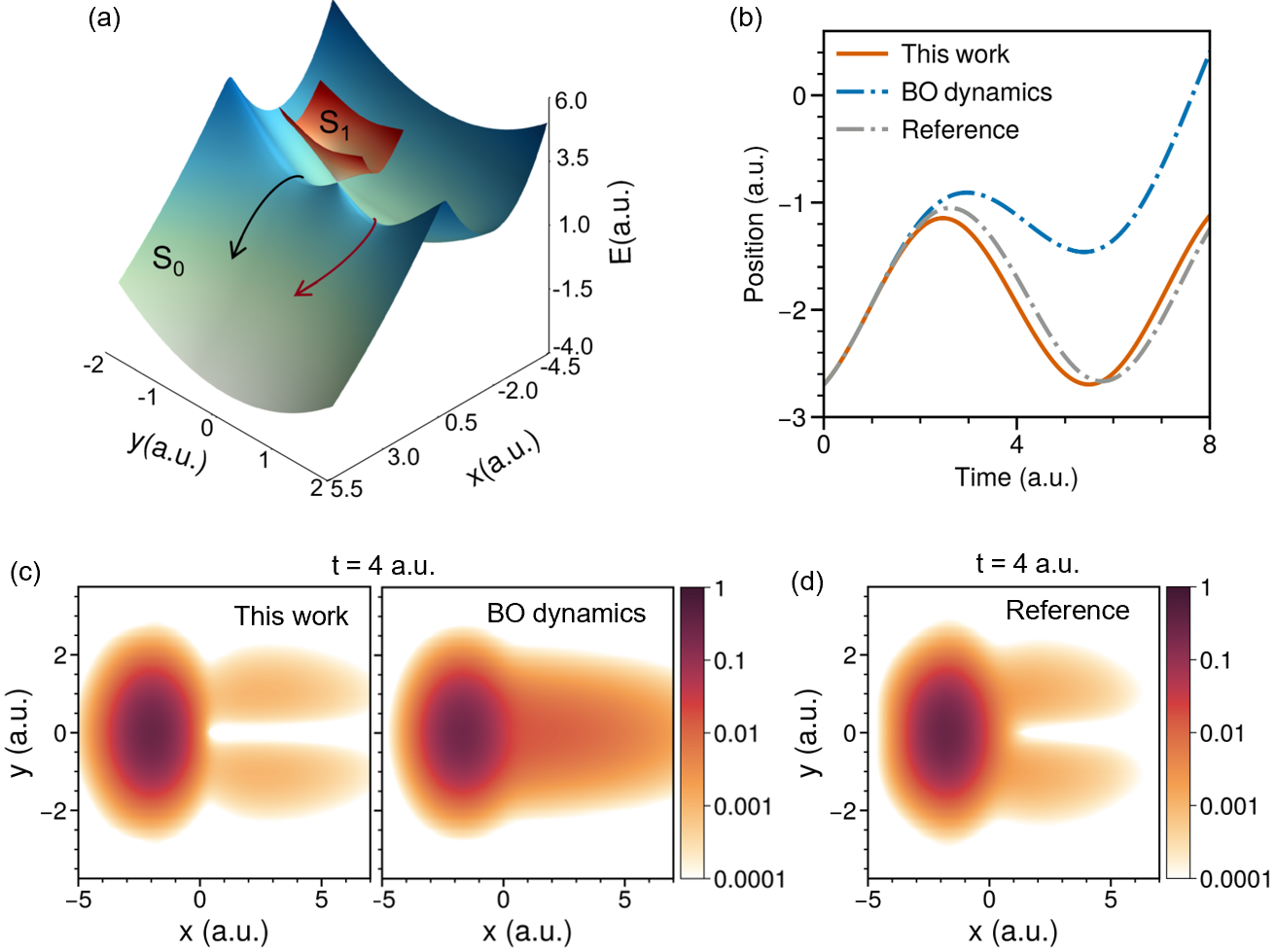}
	\caption{(a) The APESs of the vibronic coupling model. (b) The position of the nuclear wave packet. (c) and (d) display the distribution of the nuclear wave packet. The range of $x$ is $[-6, 30]$, discretized into 255 grid points, while the range of $y$ is $[-4, 4]$ with 31 grid points. The $x$ range is chosen to be large enough to prevent wave packet reflection at the grid boundaries.}
	\label{fig:model_adiabatic}
\end{figure*}

We now contrast the topological wave packet dynamics using \eq{eq:main} with BO dynamics when a conical intersection exists in an energetically inaccessible region (i.e., a barrier), ensuring that the dynamics remain adiabatic.

\subsubsection{Vibronic model}
We first consider a two-state, two-dimensional vibronic coupling model with the Hamiltonian given by
\be
H = \hat{T}_\text{N} \bf I + \bf V
\ee
where $\bf I$ is the identity matrix in electronic space, and the nuclear kinetic energy operator is defined as $\hat{T}_\text{N} =-\half \del{{\partial}^2/{\partial x^2}+{\partial}^2/{\partial y^2}}$. Here, $x$ and $y$ represent the tuning mode and coupling mode, respectively. The diabatic potential energy matrix $\bf V$ consists of two diabatic potential energy surfaces on the diagonal \cite{xie2017b}:

\be
\begin{split}
V_{11} &= \frac{\omega^2_1}{2} \del{x+\frac{a}{2}}^2 + \frac{\omega^2_2}{2} y^2 \\
V_{22} &= Ae^{-\alpha(x+b)} + \frac{\omega^2_2}{2} y^2 - \Delta
\end{split}
\ee
and the  diabatic coupling  in the off-diagonals
\be
V_{12}=V_{21}=cy e^{-(x-x_\text{CI})^2/{2\sigma^2_x} - y^2/2\sigma^2_y }
\ee
The diabatic coupling is linear around the CI and damped by a Gaussian function away from it. Here, the model parameters are set as follows: $\omega_1 = \omega_2 = 1$, $a=4$, $b=-11$, $c=2$, $A=5$, $\Delta=12$, $x_\text{CI}=0$, $\alpha=0.1$, $\sigma_x=1.699$, and $\sigma_y=0.849$.

The adiabatic potential energy surfaces (\fig{fig:model_adiabatic}a), obtained by diagonalizing the diabatic potential energy matrix, show an energetically inaccessible conical intersection flanked by two energetically lower saddle points. The CI is located at $(x, y) = (0.275, 0)$ with energy $E_\text{CI} = 2.586$ a.u. The energy of the two equivalent saddle points is 1.854 a.u., forming a potential barrier along the tuning mode ($x$) between the two wells.

We compare the adiabatic wave packet dynamics simulated with our method (which incorporates electronic quantum geometry) to the BO dynamics that assumes a trivial quantum geometry. As a reference, we also perform a simulation that includes electronically excited states, even though the dynamics remain adiabatic. All codes are implemented in our in-house Python-based package \textsc{PyQED}. The time evolution was performed using the Strang splitting method \cite{gu2024a} with $\Delta t = 0.1$ a.u.

The initial vibronic state  is a Gaussian wave packet in the electronic ground state, centered at (-2.7, 0) Bohr.
Due to the energetic inaccessibility of the conical intersection, the nuclear dynamics remains confined to the ground state.
There are two reactive pathways: one passing through the $y < 0$ shoulder and the other through the $y > 0$ shoulder, indicated by black and red arrows in \fig{fig:model_adiabatic}a.

\fig{fig:model_adiabatic}b shows the ground-state wave packet dynamics. 
Interestingly, the topological wave packet dynamics shows clearly a nodal line  along $y = 0$ (left panel in \fig{fig:model_adiabatic}c), which is a hallmark of the geometric phase effect.
This nodal structure arises because, after traversing the barrier via two reactive pathways, the two nuclear wave packets interfere destructively due to the topological phase induced by the CI, thereby creating the observed nodal line  \cite{xie2019, xie2017a, xie2016}.
This demonstrates that even when considering only a single APES and ignoring the intersecting state, our method can still account for the topological phase effect.
In contrast, when the electronic quantum geometry  is ignored by setting $A_{mn} = 1$, the wave packet dynamics (right panel in \fig{fig:model_adiabatic}c) does not exhibit a destructive quantum interference. In the diabatic model, the geometric phase can only be recovered when the full nonadiabatic model, including the electronically excited state, is simulated (\fig{fig:model_adiabatic}d).

It should be emphasized that we have only used a single ground-state potenatial energy surface. Conventionally, to incorporate geometric phase effects into nuclear motion—either through quasi-diabatization (an approximate diabatization \cite{yarkony1996a, choi2021}) or through a vector potential—it is necessary to construct the excited-state APES and first-derivative couplings \cite{xie2016, mead1982}, which can be very challenging in ab initio modeling.

\fig{fig:model_adiabatic}b displays the average position of the nuclear wave packet. Even with only the ground-state APES, the average position of $x$ obtained using our method (orange line) is in excellent agreement with the full nonadiabatic results (gray line), whereas the Born-Oppenheimer dynamics (blue line) shows a large deviation, exhibiting a faster reaction rate than it should be.
This suggests that the BO approximation can break down even for adiabatic chemical reactions, and the influence of Berry geometry can be significant.

\subsubsection{Phenol photodissociation}

\begin{figure*}[htbp] 
	\centering 
	\includegraphics[width=0.9\textwidth]{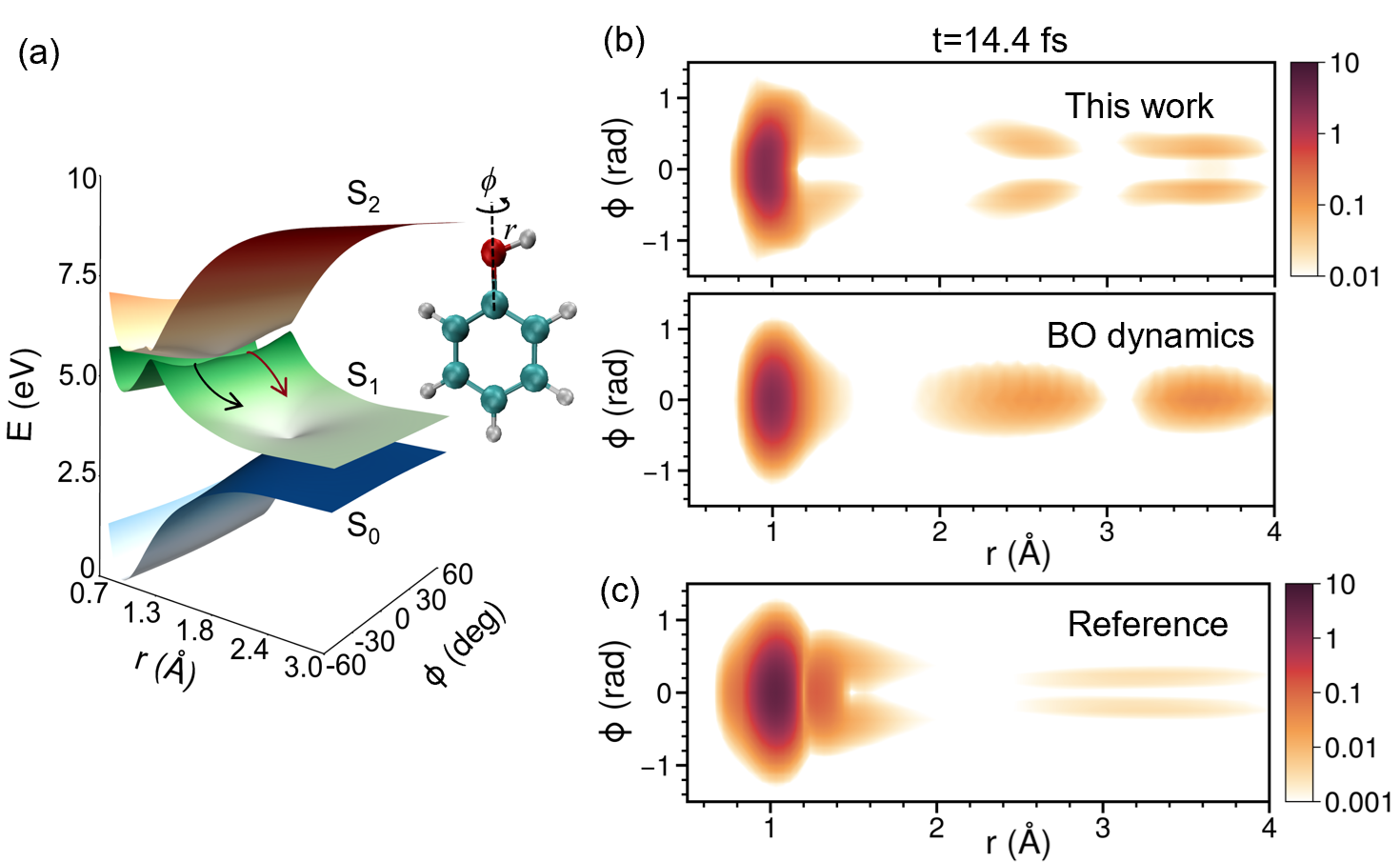} 
	\caption{(a) The geometry and APESs of phenol. (b) and (c) display the distribution of nuclear wave packets.} 
	\label{fig:phenol} 
\end{figure*}

The geometric phase in adiabatic dynamics has been demonstrated to be important in phenol photodissociation, wherein the O$-$H bond breaks via nonadiabatic tunneling \cite{zhu2016, zhu2016a, xie2017d, xu2014, yang2014}.
It has been shown that the geometric phase effect plays a critical role in this photodissociation process \cite{xie2016, malbon2016}.
The adiabatic tunneling lifetime calculated without the geometric phase effect is 0.027 ns, which is approximately 100 times shorter than the experimental lifetime \cite{xie2016}.

Here we employ the diabatic model of phenol constructed in Ref. \cite{lan2005}, which includes the three lowest diabatic states $\pi \pi$, $\pi \pi^*$, and $\pi \sigma^*$ (see Figure S1a) and two reactive coordinates $r$ and $\phi$ denoting the O$-$H bond length and the CCOH torsion angle, respectively.
The nuclear kinetic energy operator $\hat{T}_\text{N}$ in this reduced reactive coordinate space is given by
\be
\hat{T}_\text{N} =- \frac{1}{2 \mu_{\text{OH}}} \frac{\partial^2}{\partial r^2} - \frac{1}{2I} \frac{\partial^2}{\partial \phi^2}
\ee
where $\mu_{\text{OH}} = m_\text{H} m_\text{O}/ \del{m_\text{H} + m_\text{O}}$ is the reduced mass of O and H.

Direct diagonalization of the diabatic potential energy matrix produces the corresponding adiabatic potential energy surfaces (see \fig{fig:phenol}a).
The $\text{S}_1$ state intersects with the $\text{S}_2$ state at ($r=1.15$ Å, $\phi=0$) with an energy of 5.472 eV. Similar to the APESs in the previous vibronic coupling model, this conical intersection has two energy-lowering equivalent saddle points (5.391 eV) on either side, and the wave packet can cross the barrier through two distinct pathways, as indicated by the black and red arrows in \fig{fig:phenol}a.
The $\text{S}_1$ state also intersects with the $\text{S}_0$ state at a longer bond length ($r=1.97$ Å, $\phi=0$) with an energy of 3.693 eV.

The early-time dynamics of O$-$H bond cleavage occurs on the bright $\text{S}_1$ state. We focus on this early photodissociation dynamics using our topological quantum dynamics method with only a single S$_1$ state before the nuclear wavepacket reaches the S$_0$/S$_1$ CI.

The nuclear  wave functions is represented in a two-dimensional grids. The vibrational ground state is approximately described by a direct product of a one-dimensional Gaussian wave packet for the CCOH torsion angle $\phi$ and the ground state of the Morse potential for  the O$-$H stretching mode $r$
$
V_\text{Morse} = D \del{1 - e^{-a \del{r - r_\text{e}}}}^2
$, 
where $D=4.96302$ eV, $a=0.556021$ $\text{\AA}^{-1}$, and $r_\text{e}=0.9459$ $\text{\AA}$.
The range of $r$ is $(0.1, 10)$ with 127 uniform grid points, and of $\phi$ is  $(-2, 2)$ with 31 uniform grid points.
The propagator associated with the nuclear kinetic energy operators is detailed in \cref{app:kin}. 
The simulation time step is set to $\Delta t = 0.0012$ fs, with a total simulation duration of 14.4 fs. 
As confirmed in Figure S1b, the electronic population on the $\pi \pi$ state is approximately 0.4 \%  at 14.4 fs. 

The \fig{fig:phenol}b display the wave packet distribution calculated using topological wave packet dynamics and BO dynamics. 
The molecule is initially vertically excited to the $\text{S}_1$ state. The O$-$H bond stretches out and the proton comes out of the molecular plane in both directions due to the barrier at the S$_2$/S$_1$ CI. There are two reactive pathways contributing to the dissociation dynamics. 
 In our topological quantum dynamics simulation, there is clearly a nodal line along $\phi = 0$, a hallmark of geometric phase effects. By contrast,  the Born-Oppenheimer dynamics, without considering the electronic quantum geometry, does not show such destructive quantum interference. 
 To recover this effect, one has to include the S$_2$ state and perform a full nonadiabatic dynamics simulation \fig{fig:phenol}c). 
 There are slight differences in the nuclear probability distributions between our method, which considers only the $\text{S}_1$ state, and the full nonadiabatic reference calculation. This discrepancy arises because, as shown in Figure S1c, the dynamics of the model is not fully adiabatic. A small portion  ( $<$ 10\%) of the electronic population  transfers from the $\text{S}_1$ state to the $\text{S}_2$ state via the conical intersection.

\subsection{Quantum metric}

We have shown that the Berry geometry  can significantly impact nuclear dynamics in a nonlocal way, that is, even when it is not directly traversed through. We now consider the quantum metric,  encoded in the fidelity between electronic states. 

The quantum metric is a universal property in configuration space and cannot be eliminated by a gauge transformation. It becomes significant when electronic states vary drastically with molecular geometry. 
As shown below, the quantum metric can still influence the atomic motion, even in the absence of conical intersections.

\begin{figure*}[htbp]
	\centering
	\includegraphics[width=0.9\textwidth]{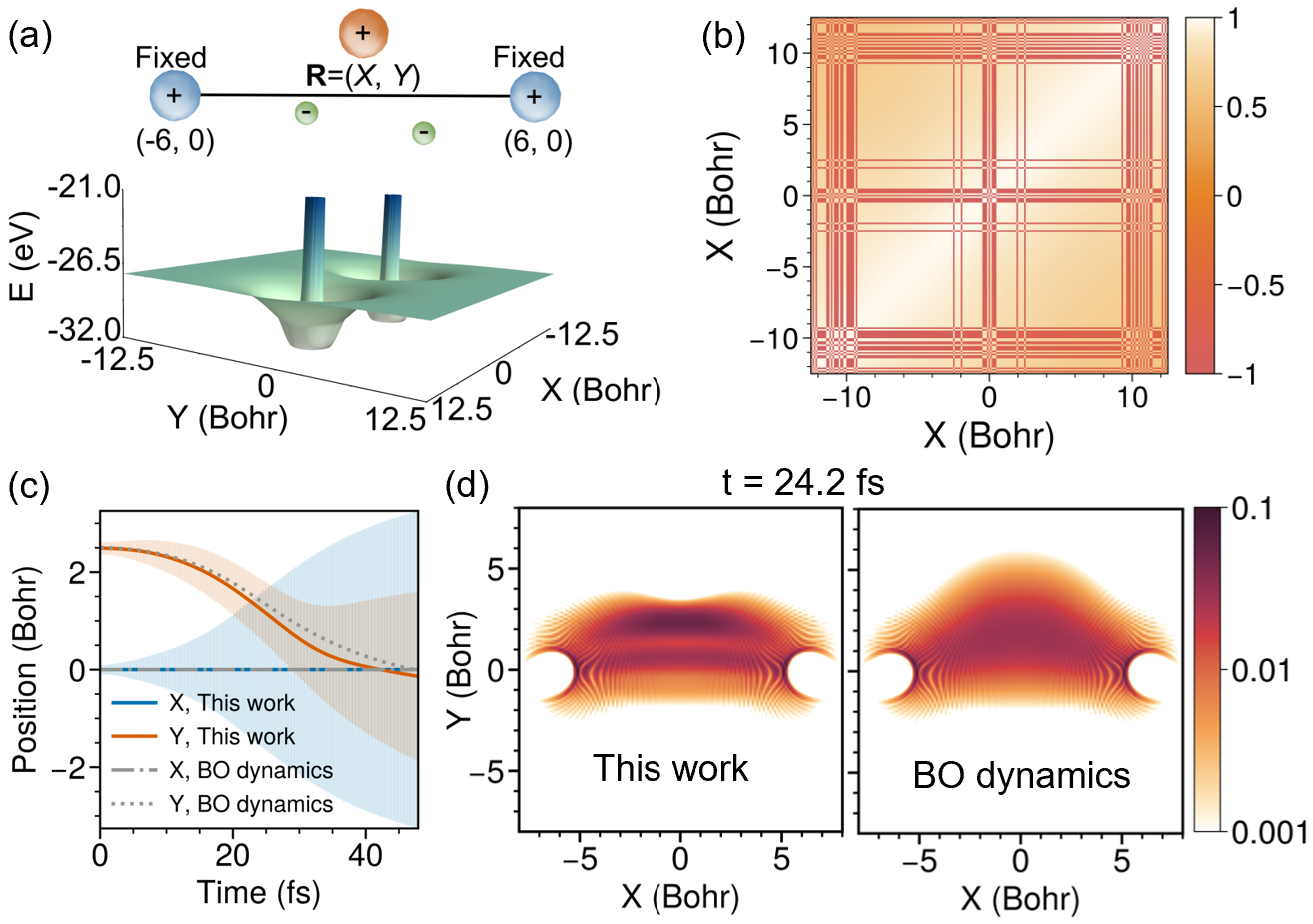}
	\caption{Adiabatic wave packet dynamics in H$_3^+$. (a) The schematic representation and ground-state APES of H$_3^+$. (b) The ground-state electronic overlap matrix of H$_3^+$ at $Y_1 = Y_2 = -12.5$ Bohr. (c) The expectation values of the position operators ($\hat{x}, \hat{y}$) computed by the two methods. (d) The wave packet distribution at 24.2 fs calculated using  our method and BO dynamics simulations.}
	\label{fig:H3+_adiabatic}
\end{figure*}

We model the adiabatic dynamics of the H$_3^+$ molecule in its electronic ground state.
We fix two protons  at $(\pm 6, 0)$ Bohr (a schematic is shown in \fig{fig:H3+_adiabatic}a), while the third proton and the two electrons are free to move within a two-dimensional plane. 
The APES (\fig{fig:H3+_adiabatic}a)  is calculated from first principles at  the level of FCI/cc-pVTZ method (full configuration interaction with the cc-pVTZ basis set) using PySCF \cite{sun2020}.
It does not exhibit conical intersections with any excited states, meaning that the effects of electronic quantum geometry arise solely from the quantum metric. At ($\pm 6$, 0) Bohr, the ground-state energy of the molecule increases sharply due to nuclear repulsion.

The global electronic overlap matrix is computed by a linked product approximation \cite{xie2025}. In it, we first compute the nearest-neighbor electronic overlap matrix elements (referred to as links $A_{n, n + \bf e_j}, j = 1, 2, \dots, d$, $\bf e_j$ is a unit vector in the $j$th direction) by directly using the many-electron configuration interaction wavefunctions.
The global electronic overlap matrix  between non-nearest-neighbor configurations are  determined by a path-ordered product of these links along a path connecting  two given geometries (see Ref. \cite{xie2025} for implementation details). The coordinate ranges for $X$ and $Y$ are both $[-12.5, 12.5]$ Bohr, with a total of 231 grid points.

\fig{fig:H3+_adiabatic}b shows the ground-state electronic overlap matrix of H$_3^+$, where the $Y$-coordinates of the two configurations corresponding to each matrix element are both -12.5 Bohr. The negative values of the matrix elements arise due to the random signs of the many-electron wave functions in the ab initio simulations. When two nuclear structures differ significantly—especially for long-range matrix elements where the two configurations are far apart—the electronic overlap deviates substantially from unity. This indicates that the quantum metric is a universal feature: even in the absence of electronic degeneracy, the electronic states still vary with nuclear geometry, consistent with our chemical intuition.

\fig{fig:H3+_adiabatic}c presents the expected proton positions, $\hat{X}$ and $\hat{Y}$ for an initial Gaussian wave packet centered at $(0, 2.5)$ Bohr. The time step is $\Delta t = 0.05$ a.u. During the dynamics, the proton gradually moves in the $-Y$ direction (see Figure S2 for the wave packet dynamics). The expectation value of $\hat{X}$ remains zero because the APES is symmetric about $X=0$, and so is the wave packet. Nevertheless, there is a large-amplitude motion of the proton in the $X$ coordinate, reflected in the variances ($\sigma_O(t) = \sqrt{\braket{O^2(t)}-\braket{O(t)}^2}$, the blue and orange shades), indicating proton delocalization in configuration space (see Fig. S2).

Despite the validity of BO approximation in this model, there is clearly a noticeable difference in the averaged $Y$ position between our method (orange line) and the BO dynamics (gray line).  The discrepancy gradually increases over time. This indicates that the quantum metric indeed influences nuclear motion. Furthermore, we observe significant differences in the wave packet distribution around [0, 3] Bohr ( \fig{fig:H3+_adiabatic}d). This difference further demonstrates that the quantum metric affects nuclear wave packet dynamics.

\begin{figure*}[htbp]
	\centering
	\includegraphics[width=\textwidth]{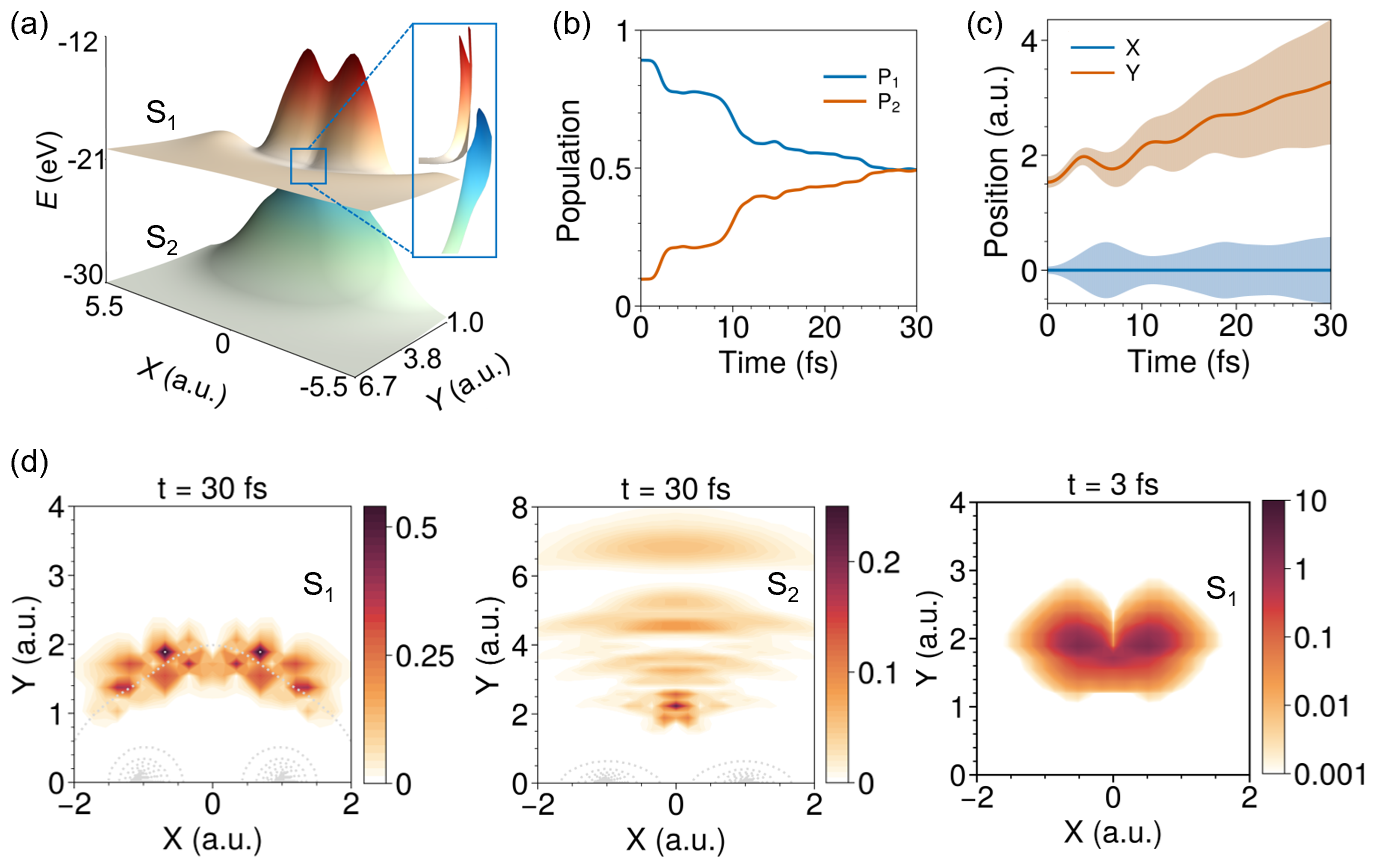}
	\caption{ Ab initio nonadiabatic  conical intersection dynamics of internal conversion in H$_3^+$.  
		(a) The APESs of H$_3^+$.  
		(b) The electronic population dynamics.  
		(c) The wave packet position of H$_3^+$.  
		(d) The wave packet distribution at 30 fs on the S$_1$ state (left) and the S$_2$ state (middle), and the distribution at 3 fs on the S$_1$ state (right).  
		The X and Y ranges are [-5.5, 5.5] Bohr and [0, 11] Bohr, respectively, with 65 grid points in each range.  
		The initial wave packet is placed at (0.0, 1.546875) Bohr. The simulation time step is $\Delta t = 0.0039$ fs.}
	\label{fig:H3+_nonadiabatic}
\end{figure*}

\subsection{Ab initio conical intersection dynamics of H$_3^+$}

To demonstrate the critical roles played by electronic quantum geometry in  nonadiabatic dynamics, we model the nonadiabatic conical intersection dynamics of internal conversion in H$_3^+$. 

Similar to before, two protons are fixed at $(\pm 1, 0)$ Bohr, while the third proton and two electrons move freely in a two-dimensional plane.
The ground and excited-state APESs and electronic eigenstates are calculated using the FCI/cc-pVTZ method. As shown in \fig{fig:H3+_nonadiabatic}a, there exists an $\text{S}_1/\text{S}_2$ CI located at (0, 1.7188) Bohr. The global multi-state electronic overlap matrix is obtained by the linked product of the nearest-neighbor electronic overlap matrices ($\bf A_{{n}, {n} + \bf e_j}, j = 1, 2, \dots, d$) \cite{xie2025}. 

 The nuclear wave packet, initially placed on the first excited state, reaches this CI at approximately 1.5 fs, triggering a rapid nonradiative electronic relaxation to the $\text{S}_2$ state. This internal conversion dynamics roughly is roughly completed within 30 fs, see \fig{fig:H3+_nonadiabatic}b for the electronic population dynamics. 
 At S$_2$ state, the proton undergoes a large amplitude motion along both directions.  
 This increasing delocalization, reflected in the variances (\fig{fig:H3+_nonadiabatic}c) is caused by the flatness of the $\text{S}_2$ APES. By contrast, as shown in \fig{fig:H3+_nonadiabatic}d left panel and also  Figure S3 upper panel, the wave packet on the $\text{S}_1$ state remains confined within the $Y$ range of 0 to 3 Bohr. 

In this internal conversion dynamics, there is a clear signature of geometric phase effects.  At 3 fs, the nuclear wave packet on the $\text{S}_1$ state forms a nodal line along $X = 0$ after passing through the conical intersection, see \fig{fig:H3+_nonadiabatic}d for the proton distribution. Analogous to the adiabatic case, this phenomenon is a result of destructive interference induced by the geometric phase effect. This simulation demonstrates that the topology of electronic states indeed captures not only nonadiabatic transitions but also geometric phase effects.

\section{Conclusion and Outlook}

By employing a discrete local trivialization of the molecular fiber bundle, we have unveiled the fundamental role that electronic geometry plays in both adiabatic and nonadiabatic molecular quantum dynamics. This approach provides an intuitive and unified quantum geometric framework for understanding effects beyond Born-Oppenheimer dynamics, including nonadiabatic electronic transitions, geometric phase effects, and diagonal Born-Oppenheimer corrections. Furthermore, our discretized ansatz for the molecular wavefunction offers a universal, divergence-free, and numerically exact approach to describing molecular quantum dynamics—both adiabatic and nonadiabatic—on geometrically nontrivial molecular fiber bundles. The challenges associated with conical intersections and, more generally, with electronic degeneracies such as non-Abelian geometric phases and exceptional points, are entirely resolved by employing the global electronic overlap matrix to capture \emph{all} effects beyond Born-Oppenheimer dynamics.

We have shown that the topological quantum molecular dynamics method provides a generic and unified conceptual and computational framework for understanding and simulating molecular quantum dynamics from first principles. For adiabatic dynamics, the global intrastate electronic overlap matrix accounts for geometric phase effects originating from energetically inaccessible conical intersections using only information of a single state (e.g., ground state), without the need to construct the vector potential, even when this potential energy surface is intersecting with other states. It also incorporates quantum metric effects. For nonadiabatic dynamics, in addition to capturing geometric phase and quantum metric effects, our approach accounts for nonadiabatic transitions due to first- and second-derivative couplings. Moreover, we have demonstrated that our method can be conveniently combined with electronic structure methods, and it immediately integrates with the extensive discrete variable representation (DVR) toolbox developed for vibrational problems (e.g., sparse grids, time-dependent DVR\cite{wang2004, lauvergnat2023b, avila2015, xie2024}). This paves the way for numerically exact \emph{ab initio} modeling of molecular quantum dynamics with strong electron-nuclear correlation.

Furthermore, we envision that the discrete local trivialization employed there is valid for all fiber bundles with complex quantum topology, including cases such as intersystem crossing, global degeneracies with non-Abelian geometric phase effects, and exceptional points in non-Hermitian dynamics.

\section{Acknowledgment}
This work is supported by the National Natural Science Foundation of China (Grant Nos. 22473090 and 92356310).

\bibliography{ALDR, dynamics, qchem, optics, topology}

\appendix

\section{Quantum Geometric Obstruction for diabatization} \label{app:dia}

In principle, transforming to the diabatic representation may remove the singular nonadiabatic couplings. 
That is to find a unitary transformation 
\be 
\ket{{\varphi}_n( \bf R)} = \sum_\alpha U_{\alpha n}(\bf R) \ket{\phi_\alpha(\bf R)} 
\label{eq:110}
\ee 
such that 
\be
\braket{\varphi_m(\bf R) | \pa_\mu \varphi_n(\bf R)} = 0 
\label{eq:130}
\ee 
The diabatic states will not be eigenstates of the electronic Hamiltonian, and do not vary significantly with nuclear geometries. 

Inserting \eq{eq:110} into \eq{eq:130} leads to 
\be
\pa_\mu U_{\beta n}(\bf R) + \sum_\alpha F^\mu_{\beta \alpha}(\bf R) U_{\alpha n}(\bf R) =0
\label{eq:113}
\ee 
If \eq{eq:113} admits a solution,  this leads to the necessary condition  
\be
\pa_\mu \bf F_\nu - \pa_\nu \bf F_\mu + [\bf F_\mu, \bf F_\nu] =  0
\label{eq:na}
\ee  

The left-hand side of \eq{eq:na} can be shown to be precisely the antisymmetric component of the non-Abelian electronic quantum geometric tensor 
\be
\bm \Omega_{\mu \nu} = [\bf D_\mu, \bf D_\nu]   = \bf Q_{\mu\nu} - \bf Q_{\nu \mu }. 
\ee 
where $\bf D_\mu = \pa_\mu \bf I + \bf F_\mu$ is the covariant  matrix derivative. 
The condition for diabatization  is thus  
\be 
\bm \Omega_{\mu \nu} = 0 
\ee  
This condition cannot be satisfied. 
Not only the the non-Abelian Berry curvature matrix does  not  vanish for general polyatomic systems \cite{kendrick2002, mead1982} but it is typically not a small number.

\section{Derivation of \eq{eq:111}} \label{app:bh}
Inserting the Born-Huang expansion \cref{eq:bh} to the molecular \tdse  
\be 
i \pd{\Psi(\bf r, \bf R, t)}{t} = \del{- \frac{1}{2M_\mu} \pa_\mu^2 + H_\text{BO}(\bf r; \bf R) }\Psi(\bf r, \bf R, t) 
\ee 
yields 


\begin{align}
i \phi_a \pd{ \chi_a}{t} = & - \frac{1}{2M_\mu} \del{ \del{\pa^2_\mu \phi_\alpha} \chi_\alpha + 2\pa_\mu \phi_a \pa_\mu \chi_a + \phi_a \pa_\mu^2 \chi_a} \nonumber \\
& + V_\alpha(\bf R) \phi_a \chi_a 
\end{align}

where $V_\alpha(\bf R)$ is the $\alpha$th adiabatic potential energy surface. 
Left-multiplying $\phi_\beta(\bf r; \bf R)$ and integrating over electronic coordinates yields 

\begin{align}
i \pd{ \chi_\beta(\bf R, t)}{t} = & \del{ -\frac{1}{2M_\mu} \pa_\mu^2 + V_\beta } \chi_\beta(\bf R, t) \nonumber \\
&  - \frac{1}{2 M_\mu} \sum_{\alpha} G_{\beta \alpha} \chi_a - \frac{1}{M_\mu} F_\mu^{\beta \alpha} \pa_\mu \chi_a   
\label{eq:112}
\end{align}
Making use of the identity
\be \pa_\mu F^{\beta \alpha}_\mu = G^{\beta \alpha}_\mu + \braket{\pa_\mu \phi_\beta }{ \pa_\mu \phi_\alpha}, \ee 
\eq{eq:112} can be rewritten as 

\begin{align}
i  \pd{ \chi_\beta(\bf R, t)}{t} = & { -\frac{1}{2M_\mu} (\pa_\mu + F_\mu)^2 + V_\beta } \chi_\beta(\bf R, t) \nonumber \\
& + \frac{1}{2M_\mu} \braket{\pa_\mu \phi_\beta }{ \pa_\mu \phi_\alpha}   + \frac{1}{2M_\mu} \bf F_\mu \bf F_\mu 
\label{eq:126}
\end{align}
Using the antisymmetry of the first-derivative coupling 
\be F_\mu^{\beta \alpha} = - \overline{ F^\mu_{\alpha \beta}} \ee in \cref{eq:126}
yields \eq{eq:111}.

\section{Kinetic energy propagator in Jacobi coordinates} \label{app:kin}

 For Jacobi coordinates $(r, \theta)$,  the kinetic energy operator reads 
 \be 
 \hat{T}_\text{N} = \frac{p_r^2}{2\mu} + \frac{p_\theta^2}{2I(r)} 
 \ee 
 where $\mu$ is the effective mass for the stretching mode and $I(r)$ is the moment of inertia. 

 In the Trotter decomposition of the short-time propagator \cite{gu2024a}, the propagator associated with the kinetic energy operator reads 
 \be 
 e^{-i \hat{T}_\text{N} \Delta t} =  e^{-i \hat{T}_r  \Delta t} e^{-i \hat{T}_\theta \Delta t} + \mc{O}(\Delta t^2)
 \ee 
 In the Jacobi coordinates, the two kinetic energy operators do not commute as the moment of inertia depends on the bond length.
 
 The kinetic energy operator for the stretching coordinate can be computed easily by DVR. 
 For the angle, 
we compute the kinetic energy matrix $\bf T_\theta$  first and then take a exponentiation $e^{-i \bf T_\theta \Delta t}$
 The matrix elements read 

\be 
\mel{r_{n'} \theta_{m'} }{ \frac{p_\theta^2}{2I( r)}  }{ r_n \theta_m} = \delta_{n'n} \frac{1}{2I(r_n)} \mel{\theta_{m'} }{ p^2_\theta }{\theta}  
\ee 
The matrix elements for the kinetic energy propagator read 
\be 
\mel{r_{n'} \theta_{m'} }{ e^{- i  \frac{  p_\theta^2   }{2I(\bf r)} \Delta t }  }{ r_n \theta_m} = \delta_{n'n}  \mel{\theta_{m'} }{ \exp\qty{ - i \frac{p^2_\theta}{2I_n} \Delta t} }{\theta_m}  
\ee 
where $I_n \equiv I(r_n)$.
Thus, 
\be 
e^{-i \hat{T}_\text{N} \Delta t} \ket{\bf n \alpha} \approx \sum_{\bf m, \beta}  A^{\bf m \beta}_{\bf n\alpha} \mc{K}^r_{m_1 n_1} \mc{K}^\theta_{n_1, m_2 n_2}  \ket{\bf m \beta}
\ee 
where 
\be 
\mc{K}^\theta_{n_1, m_2 n_2} = \mel{n_1, m_2 }{ e^{-i \hat{T}_\theta \Delta t} }{ n_1, n_2}
\ee 

%
%
%
%
%
%

\end{document}